\documentclass{ewic-latex}
\usepackage{graphicx}
\usepackage{booktabs}
\usepackage{hyperref}
\usepackage{url}

\begin{document}

\runningheads{Anonymous}{Computer Science Education \& GenAI}

\conference{Proceedings of \dots}

\title{Computer Science Education in the Age of Generative AI}

\authorone{Russell Beale}

\begin{abstract}
Generative AI tools—--most notably large language models (LLMs) like ChatGPT and Codex—--are rapidly revolutionizing computer science education. These tools can generate, debug, and explain code, thereby transforming the landscape of programming instruction. This paper examines the profound opportunities that AI offers for enhancing computer science education in general, from coding assistance to fostering innovative pedagogical practices and streamlining assessments. At the same time, it highlights challenges including academic integrity concerns, the risk of over-reliance on AI, and difficulties in verifying originality. We discuss what computer science educators should teach in the AI era, how to best integrate these technologies into curricula, and the best practices for assessing student learning in an environment where AI can generate code, prototypes and user feedback. Finally, we propose a set of policy recommendations designed to harness the potential of generative AI while preserving the integrity and rigour of computer science education. Empirical data and emerging studies are used throughout to support our arguments.

\end{abstract}

\maketitle
\section{Introduction}
The advent of generative artificial intelligence has generated a seismic shift in many fields, and computer science education is no exception. Since the public release of ChatGPT and similar models, educators and students alike have witnessed a dramatic change in how programming can be taught and learned. LLMs such as OpenAI’s Codex are capable not only of generating coherent text but also of writing code in multiple programming languages with surprising fluency. This capability presents both immense opportunities and significant challenges for computer science (CS) education.

\subsection{Background and Rationale}
Traditionally, computer science education has focused on teaching students the fundamentals of algorithms, data structures, programming languages, and software engineering practices. Students are expected to learn not only to write code but also to develop debugging skills, algorithmic reasoning, and problem‐solving techniques. they learn HCI skills to help them identify problems, learn techniques to produce creative options, and prototype and evaluate solutions.  However, with the rise of AI-powered code assistants, the very nature of coding and prototyping is changing. Today, an LLM can generate code snippets, provide explanations for errors, create prototype flows and interactive demos, and even suggest complete solutions to  tasks.  Given that AI can generate code on demand, traditional programming assignments are no longer be a reliable indicator of a student’s understanding: the link between code production and understanding is broken.  In HCI, developing artifacts can be done using GenAI for many of the stages, and so we have to rethink the educational priorities and agenda.

This technological evolution calls for a radical rethinking of CS education. Educators must now decide:
\begin{itemize}
\item What is the fundamental aim of teaching programming?
\item What are the key skills we want Computer Science graduates to have?
\item What should be taught?
\item Should the curriculum focus more on understanding, verifying, and improving AI-generated code/examples rather than on writing code from scratch?
\item How should it be taught?
\item How can instructors leverage AI tools to foster active learning, creativity, and critical thinking?
\item How should work be assessed?
\end{itemize}

\subsection{Objectives of This Paper}
In this paper, we address the following key questions:
\begin{itemize}
\item \textit{Opportunities:} What are the promising applications of generative AI in computer science education? We examine how AI can enhance coding assistance, debugging, personalized feedback, and overall learning efficiency.
\item \textit{Challenges:} What risks do these technologies pose?   We explore issues such as academic dishonesty, over-reliance on AI, and the inherent difficulties in assessing originality.
\item \textit{Curricular Implications: }What should be taught in CS programs in light of these new tools?  We discuss the need for integrating code literacy, critical evaluation of AI output, and ethical use of AI into the curriculum.
\item \textit{Pedagogical Strategies:} How should educators modify their teaching methods? We propose strategies such as AI-augmented pair programming, active learning, and process-based assignments.
\item \textit{Assessment Approaches:} How can we design assessments that measure genuine understanding and creativity?  We consider alternative assessment models such as oral examinations, code walkthroughs, and iterative project submissions.
\item \textit{Policy Recommendations:} What institutional policies are needed to manage the dual-edged nature of generative AI in CS education?
\end{itemize}

We structure our discussion across these themes, drawing on empirical studies, recent surveys, and case studies from both US and UK institutions. Although the literature specifically targeting the use of AI in CS education is still emerging, we rely on a combination of general studies on generative AI in higher education and research focused on AI and code generation.

\section{Opportunities: Harnessing AI in Computer Science Education}

Generative AI has opened new horizons in how computer science can be taught and learned. In this section, we explore the key opportunities that LLMs offer for CS education.

\subsection{Enhancing Code Generation and Debugging}
\subsubsection{AI as a Code Assistant}

One of the most significant developments in recent years is the ability of AI to generate code. In a landmark study, \citet{Chen2021} evaluated large language models trained on code and demonstrated that these models can produce syntactically correct and functionally useful code across multiple programming languages. This capability is transforming the role of the programmer from writing code line-by-line to orchestrating and refining AI-generated code.  Discussions on forums such as Reddit amongst technical people are around which sorts of models are best for which sorts of tasks: this shows both that they are not perfect at coding yet, but mostly that the conversation has completely moved on from "can I use..." to "what's best to use?" 

Many modern IDE's provide close integration with LLMs, making them accessible directly from the codebase, mouse menus and information panels, so that using them is much like using code completion used to be.  For students, having access to AI-powered code assistants (such as GitHub Copilot) can reduce the frustration often associated with early programming assignments. For example, \citet{ma2024integratingaitutorsprogramming} report that 78\% of students using the tutor reported it helped their learning.  These tools provide immediate feedback, suggest fixes for errors, and help clarify programming concepts by generating working examples. As a result, students can focus more on understanding the underlying algorithms and less on memorizing syntax.  A semester long study \citep{lyu_evaluating_2024} evaluated how novices use LLMs, demonstrating a significant significant improvement in their scores.  In 10-17 year olds learning to program\citep{kazemitabaar_how_2024}, a hybrid approach with students collaborating with the AI tools showed the best outcomes for the students.

\subsubsection{Debugging and Code Optimization}
In addition to generating code, LLMs can assist with debugging and code optimization. When students encounter errors, AI can help pinpoint the cause of the bug and propose corrections \citep{lee_code_2024, pdurean_bugspotter_2025}. This can be particularly valuable in introductory courses, where debugging is a critical—yet challenging—skill\citep{ma_how_2024}. . By using AI as an on-demand tutor for debugging, students can learn how to diagnose and fix issues more efficiently Usage is simple: Bugs are literally highlighted for the LLM which then provides guidance as to the likely source of error, and will rewrite the relevant code segment to correct it.  Conceptual errors, where the code doesn't do what is expected, can be highlighted and explained by the student, and again the code will be refined and re-written.  Similar approaches apply for edge cases, and so on.

Moreover, advanced AI tools can analyze code for efficiency and suggest optimizations - for example Gao et al. demonstrated a more than doubling of execution speed using their LLM approach\citep{gao_search-based_2024} . This not only reinforces good programming practices but also encourages students to think critically about code performance, resource management, and algorithmic complexity.

\subsection{HCI Education}
If we consider the user-centred design cycle of empathise-define-ideate-prototype-test, GenAI can assist in all stages: it can be tasked to act as a client to help understand the problem, identify usability problems \citep{graser_using_2023}, can simulate user behaviours \citep{wang_user_2025},  provide user research or act as a persona \citep{schuller2024generating, sun2024persona}, can create prototype workflows and create interactive systems from them \citep{vercel_v0_2025}, and run usability tests on the results \citep{pourasadDoesGenAIMake2025}. These remarkable capabilities allow students to much more fully explore possibilities than ever before, especially within the confines of a three month course.  The ability to much more rapidly understand the domain, explore options and develop testable prototypes means that the skills previously required for these taks are much more accessible to all, and so the focus shifts to real empathy and understanding in the mind of the student.

\subsection{Personalized Learning and Adaptive Feedback}
\subsubsection{Tailored Instructional Content}

Generative AI can customize instructional materials to meet individual student needs\citep{abolnejadian_leveraging_2024}. By analyzing a student’s performance data and learning style, AI systems can generate adaptive content \citep{hou_codetailor_2024}—such as customized examples, targeted practice problems, and explanations in varying levels of complexity. This adaptive learning approach ensures that each student receives the support they need, thereby improving overall engagement and learning outcomes.

\subsubsection{Adaptive Feedback on Programming Assignments}
One of the most promising applications of AI in CS education is the provision of adaptive feedback. \citet{Kinder2024} demonstrated that AI-generated feedback can significantly improve the quality of student submissions by highlighting areas for improvement and suggesting specific modifications. In programming courses, this means that instead of simply marking an assignment as “incorrect,” an AI tutor can provide detailed commentary on why a particular solution fails and how it might be improved. Such feedback not only aids in the immediate learning process but also encourages students to adopt a more reflective approach to problem-solving.

\subsection{Facilitating Collaborative and Peer Learning}
\subsubsection{AI-Augmented Pair Programming}

The traditional model of pair programming\citep{hanks_pair_2011, begel_pair_2008}—where two students work together at one workstation—has been shown to improve learning outcomes by fostering collaboration and knowledge sharing. With the advent of generative AI, a new variant known as “AI-augmented pair programming” is emerging\citep{moradi_dakhel_github_2023, bird_taking_2023}. In this approach, the AI acts as the  partner in the programming process, offering suggestions and corrections \citep{ma_is_2023}, and are just as trusted and respected as a human partner\citep{kuttal_trade-offs_2021}. This model not only enhances the collaborative experience but also introduces students to modern software development practices where AI tools are becoming standard.

\subsubsection{Community and Forum Support}
Online programming communities, forums, and collaborative coding platforms have long been part of the CS educational landscape. Integrating AI into these platforms can further support peer learning. For example, AI can moderate forums, answer frequently asked questions, and even summarize long threads of discussion, thereby making community knowledge more accessible. Such tools can help maintain a high level of engagement in online courses and ensure that students receive timely support outside of class hours.

\section{Challenges and Risks: The Dark Side of AI in CS Education}

While generative AI offers exciting opportunities, it also introduces significant challenges. In the context of computer science education, the risks often revolve around the potential for superficial learning, over-reliance on automated solutions, and academic integrity.

\subsection{Pedagogical Concerns}
\subsubsection{Superficial Learning and Skill Dilution}

Generative AI may encourage students to take shortcuts. Instead of engaging in the iterative process of writing, testing, and debugging code, students might simply accept the AI-generated output as correct. This risk is particularly acute in introductory courses, where mastering the fundamentals of programming is essential. The educational goal of developing a deep understanding of algorithmic processes and programming logic may be compromised if students bypass these critical learning experiences. One survey highlights the issues, going as far as to say that it is "premature to incorporate these systems into computing education" \citep{pirzado_navigating_2024} though many others disagree \citep{kazemitabaar_how_2024, kazemitabaar_codeaid_2024, lyu_evaluating_2024}.  The general consensus appears to be that well-structured material that engage with LLMs as programming partners and encourage analysis of the produced code, iterative debugging and problem refinement, lead to better outcomes and deeper understanding.

The same is true for other topics in computing education, especially in HCI.  Most aspects of interactive design activities can be outsourced to GenAI to do a passable job, and this can lead to superficial understanding from the student as they have not engaged in progressive problem solving and the 'useful struggle' that is essential to experiential learning.

\subsubsection{Ethical and Trust Issues}

The integration of AI into the classroom raises complex ethical questions. For instance, if AI-generated outputs are not properly disclosed or if its use is not appropriately monitored, the integrity of the educational process is jeopardized. Moreover, there is a risk that students might develop a misplaced trust in AI outputs, assuming that the results generated are always correct or optimal. This can lead to errors that propagate into larger projects and ultimately undermine the learning process. Educators must, therefore, emphasize the importance of critical evaluation and independent verification of AI-generated work.

\subsection{Academic Integrity and Plagiarism}
\subsubsection{Over-Reliance on AI-Generated Artifacts}

One of the most pressing concerns is that students may rely too heavily on AI to generate artifacts for assignments (whether code, prototypes, essays, or data), leading to a situation where the submitted work no longer reflects the student’s own understanding. \citet{Paustian2024} found that nearly 47\% of students in a study admitted to using LLMs in their coursework, with 39\% using them for exam or quiz questions and 7\% for entire assignments. Although these figures come from a broader study, similar trends are emerging in computer science courses where code generation is directly at stake.  A recent informal evaluation in a functional programming course for which AI use had been prohibited, when explicitly asked nearly 30\% admitting using it, with likely 50\% actual usage.

Over-reliance on AI tools may lead to a decline in the development of fundamental programming skills. When students can simply “prompt” an AI for a solution, they may not fully engage with the problem-solving process, hindering the development of critical thinking and debugging abilities.  However, concerns about this may be overblown.  A survey by  \citet{chan_students_2023} on student perceptions of GenAi is education found that "Another reason behind the optimism was the assumption that humans would still maintain control and oversight over the GenAI" suggesting that they are not just passive recipients of what it tells them.  There is little evidence on how much students are likely to blindly present the results of GenAI solutions to problems with no further checking:  a small scale study on 15 students by \citet{amoozadeh_student-ai_2024} found one third of students simply submitted the whole task to ChatGPT, and that few students verified their solutions.  If  students simply accept AI-generated solutions, they risk bypassing important learning stages such as debugging, error analysis, and algorithmic reasoning.  Other studies lke the 455-student programming homework help evaluation by \citet{ma2024integratingaitutorsprogramming} are more positive: about half the students engaged with the tutors and 78\% of these reported it helping them. Conversely, when integrated properly into the curriculum, students can use AI as a starting point. In such cases, they review, debug, and iteratively improve on the generated code. This process can reinforce their understanding by prompting them to critically assess the AI’s output. For instance, the earlier work of  \citet{Allamanis2018} show that using AI as a collaborative tool encourages deeper analysis and comprehension.  The key factor is the instructional design. If educators build assignments and assessments that require students to explain, critique, and modify the AI output, then the use of LLMs can indeed promote deeper learning rather than just offering a shortcut. However, without such scaffolding, there is a potential risk that students may bypass essential learning processes. In summary, the concern is real if AI tools are used uncritically. But with effective pedagogical strategies—such as requiring code walkthroughs, debugging sessions, and reflective write-ups—LLMs can serve as a powerful aid in developing coding skills.

\subsubsection{Enforcement and detection: the wrong questions}
Most authors argue for strong enforcement measures to ensure academic integrity.  Whilst there is undoubtedly a place for these in the very short term, this seems to be the wrong issue to focus on.  Detection of LLM use across natural language subjects varies from maybe 88\% \citep{Paustian2024}- but code plagiarism is particularly hard; \citet{Chen2021} show code can often be syntactically correct and functionally viable, making it harder to distinguish from human-written code solely by surface-level features.  As LLM code generation capabilities increase, the ability for tools to detect their influence is likely to wane.  But more critically, it suggests that we want to test how well students can write code without assistance, which seems a less sensible choice.  Is it better to ask students to write code on paper, or in an editor? Do we care if the system autocorrects their spelling?  A better approach is to refine assessments and learning challenges so that the requisite skills are obtained through the process, rather than relying on inspecting the outcome to assess it. 

\subsection{Disciplinary Disparities and Access Issues}
\subsubsection{Variability Across Courses and Levels}

In computer science education, the impact of generative AI may vary significantly depending on the course and the level of study. While advanced courses might incorporate AI tools as part of a broader suite of resources, introductory courses face the challenge of ensuring that students develop basic  competencies without over-reliance on AI. Tools such as CodeAid \citep{kazemitabaar_codeaid_2024} aim to provide LLM stupport without providing direct answers, a sensible integration of pedagogical strategy into LLM capability. 

\subsubsection{Socioeconomic and Gender Considerations}

Empirical surveys, such as the HEPI (2025) report\citep{HEPI2025}, have indicated that demographic factors (including socioeconomic status and gender) can influence the adoption of AI tools. In computer science, where there is already a noted gender imbalance, the risk exists that AI tools may inadvertently widen the gap if not all students have equal access or the same level of digital literacy. However, \citet{hanks_pair_2011} provide evidence that women may particularly benefit from AI pair programming.  Ensuring equitable access to AI resources—and incorporating training to bridge these gaps—is essential for maintaining fairness and inclusivity in CS education.

\section{What Should We Teach in the AI Era?}

The rise of generative AI necessitates a rethinking of computer science curricula. Educators must decide not only which traditional topics remain essential but also how to incorporate new skills that are critical in an AI-augmented landscape. The general tone of previous articles in this space has been that there is a strong need to ensure that the fundamentals of syntax and semantics, data structures, complexity and the ilk are necessary skills to be taught within the curriculum, but it is worth questioning this approach (which is essentially the status quo).

No-one teaches programming in binary.  Assembler programming is becoming rarer and rarer.  Programming languages are trending towards more human-readable ones (Python, Ruby rather than Fortran, C, for example).  As the tools available to us have increased in capability, they have also increased our abstraction from the underlying hardware - and many of us would consider that A Good Thing.  It could be argued that we are on the brink of another shift in the default level of programming we should consider.  If we look at how some thought leaders are playing with the technology, we see examples of "vibe coding" \citep{karpathy_theres_2025} - Karpathy - Senior Director of AI at Tesla, and a founding member of OpenAI - discusses how coding has become a conversation (literally, he uses voice recognition software) with the GenAI system to play with ideas, refining, exploring pathways, and generally exploring concepts interactively.  I'm pretty sure he knows the basics of syntax, semantics and data structures - but he's not thinking about them when problem solving.

Or look at how people are using LLMs for many tasks that they might previously have had to program for. A personal example - I had written some code to help my own research activities which, when given an article, would parse it,, extract the references from that paper, perform an appropriate search to locate those references and summarise them, building a network graph of the relationship between them and their topics.  Whilst it worked, it often had to be tweaked and fixed: some articles were hard to parse; ensuring we abided by robotic access terms to websites and papers was ever-changing; cookies caused issues.  Now, I can ask ChatGPT to do this, and can refine what it looks for and how it accesses it by refining my prompts and interactions, threading my history to clarify what contexts should be used when.  Thus prompt engineering\citep{phoenix2024prompt, marvin_prompt_2024}, as it is now called, is a critical tool in my arsenal, and has superseded my own tool.  This is, by my definition, programming.  What is interesting is that it requires similar skills to standard code programming to work effectively: problem solving, abstraction and refinement, focusing on small areas to correct.  It can also be tackled in similar ways to coding. For code, one can either quickly create the core things that work for much of the problem, and then refactor and refine it, segmenting it not more manageable chunks and checking they work on all edge cases as well as conventional ones.  Or one can devise the overall architecture and structure initially, and then write code fragments and test suites that provide the specific functionalities for each part.  Depending on the size of the project, one approach or another is preferred.  And so it is with LLMs - you can ask a general question  and then pursue the lines of enquiry that ensure, refining and developing understanding as you go,  or you can give it a structure and ask for details on specific things, collating and linking themes.  We are now starting to program in English, not Java or Python or C++: conversational programming is here to stay\citep{esposito2024programming}.  This makes a particularly strong case for replacing coding teaching for non-programmers, non-Computer Scientists with courses on prompt engineering, iterative refinement of conversational explorations of topics, and thus giving significant power and skills to less technically-minded people. However, this paper is not about that: it is about what Computer Science education should look like.

It seems fundamental that first year Computer Scientists understand code structure, data structures, and algorithms, and should understand what syntax and semantics mean.  This is because some of these people will be going on to create low level technologies that need awareness of these issues; most, however, do not. We generally teach them enough about binary and hardware architecture, and compilers, for them to understand the pipeline from code to execution.  Given this perspective on these lower level technologies, it is appropriate to ask if some of these other basics have moved to a similar category.  It seems reasonable to allow students to move from the current approach that requires recall of these concepts and move instead towards recognition. Therefore, we propose a curriculum that doesn't explicitly test detailed knowledge of  syntax and semantics, for example by requiring students to write code in paper under exam conditions, but rather to be able to play with code on screen and recognise when it's correctly formed and when it isn't.  By exploring, they can identify what a loop is doing, how it increments and exits; they can understand lists and sorting.  For many of the higher order data structures, libraries exist that use optimised code to perform the relevant functionality, and teaching awareness and use of these libraries is a core part of current programming activity.  Conceptually, it's important to understand what adding items into a list means - and that can be taught with pieces of paper on a desk; being able to write characters on a screen in order to achieve it adds little to the comprehension process. 

\citet{dwivedi_so_2023} in "So What if ChatGPT wrote it?" provide a multidisciplinary perspective on the problem and settle on the usual benefits and issues, but the issue is more complex around learning to code, because, after all, code is either right, or wrong.  Some might argue that 'compiles but doesn't fully work' is an intermediate stage - nevertheless, there are limited ways to solve the sorts of problems set in programming classes, and quite which entity puts the letters into the IDE is perhaps not the biggest issue.  But if users see what's there, play with it, dissect it and reassemble it, then they get a decent comprehension of the features.  Does it matter if it was written by ChatGPT, if they can explain it properly, identify common data structures and functionality, and describe why the code is structured this way rather than another? I don't think so.

Thus, I am strongly advocating exploration and iteration - just seeing what GenAI puts on the screen doesn't really help - but changing it, noticing the errors, identifying edge cases and refining it, all give practice in using appropriate code and recognising when it's correct. Seeing lots of syntactically correct code is also a positive; it's quite feasible that many students struggle through assignments until their code compiles and the first  correct version they see is the final one they submit.  As an analogy, if we want to teach someone to make meringue, learning about how eggs are made is not much use - instructions for whisking are helpful, but really it's only by doing it a few times and then becoming aware of what textures to look for does one become any good at creating them.

\subsection{Integrating AI Literacy}
Given that AI tools are now part of the software development landscape, students must learn how to use them ethically and effectively. Key topics to incorporate include:
\begin{itemize}
\item \textit{AI-Augmented Development Workflows:} Practical lessons on integrating AI tools (such as GitHub Copilot or Codex) into software development workflows can help students learn modern programming practices. This includes learning how to use version control systems alongside AI tools and developing strategies for collaboration with AI.
\item \textit{Understanding LLMs and Their Limitations:} Students should be introduced to the basic principles of how LLMs work, including the concepts of training data, model bias, and the phenomenon of “hallucination” (i.e., generating plausible but false outputs)\citep{huang_survey_2025}.
\item \textit{Critical Evaluation of AI-Generated Code:} Coursework should emphasize that AI-generated code is a starting point rather than a final product. Students must learn to review, test, and improve on the AI output.
\item \textit{Ethical Considerations:} Discussions on academic integrity, plagiarism, and the responsible use of AI should be integrated into the curriculum. Students need to understand both the benefits and the potential risks associated with AI assistance.
\end{itemize}

\section{Pedagogical Strategies: How Should We Teach in the AI Era?}

The transformation brought by generative AI calls for innovative teaching strategies that not only integrate new tools but also preserve the rigour of computer science education. In this section, we discuss several pedagogical approaches that can help achieve these goals.

\subsection{Incorporating AI Tools into the Classroom}
\subsubsection{Demonstrations and Interactive Labs}

One effective approach is to integrate live demonstrations of AI-powered coding tools into lectures and lab sessions. Instructors can show how to use tools like GitHub Copilot for generating code, debugging, and optimizing algorithms. Interactive labs can be structured so that students work in small groups to solve problems using both traditional methods and AI assistance. This dual approach allows students to compare outcomes, understand the limitations of AI, and develop a balanced perspective on its role in software development.

In HCI, we can create a series of personas from data collected in exercises, and interact with them in depth to gain a deep understanding of their needs and pain points in a way unmatched by paper-based descriptinos.

\subsubsection{AI-Augmented Pair Programming}

Pair programming has long been an effective pedagogical strategy in computer science. With AI-augmented pair programming, the classroom dynamic shifts to include the AI as the partner, or as a third partner. Students work in collaboration with the AI system that offers suggestions, explains errors, and provides alternative approaches. This method not only improves collaboration but also familiarizes students with modern development workflows where human–AI interaction is increasingly common.

\subsubsection{Flipped Classroom Models}

In a flipped classroom model, students engage with lecture material (including tutorials on using AI tools) before coming to class. Class time is then devoted to problem-solving sessions and discussions. Instructors can design activities that require students to critique AI-generated solutions or improve upon it, thereby fostering deeper engagement with the material. In programming classes, alternatives include providing code with deliberately introduced bugs \citep{pdurean_bugspotter_2025, ma_how_2024}, and asking students to fix the code. Such an approach encourages active learning and critical thinking.  In HCI situations, poor interaction flows can be presented and critiqued by the students working in parallel with AI to get persona, usability and new alternatives to explore different perspectives.

\subsection{Emphasizing Process and Critical Thinking}
\subsubsection{Artifact Review and Walkthrough Sessions}

Rather than simply grading the final  output, educators can emphasize the importance of the development process. Artifact review sessions—where students present their outputs and explain their design decisions—help instructors assess a student’s understanding. These sessions also provide an opportunity for peer feedback, where students learn to critique and improve on both human- and AI-generated solutions.

\subsubsection{Project-Based Learning and Iterative Development}

Project-based learning (PBL) remains a powerful way to teach. In the context of AI, PBL can be designed to require multiple iterations. Students can submit preliminary versions of their projects, receive AI-generated as well as human feedback, and then revise their work. This iterative process emphasizes the importance of refinement, testing, and validation—skills that are crucial when working with AI-generated solutions.  This approach works as well for HCI design projects as it does for coding exercises, or for database designs or full stack interactions.

\subsubsection{Emphasizing Reflective Practices}

To counteract the risk of superficial learning, educators should incorporate reflective practices into the curriculum. For instance, students might be required to write reflective essays on the role of AI in their development process, discussing both its benefits and limitations. This reflective component encourages a meta-cognitive approach, ensuring that students are not merely accepting AI outputs at face value.

\subsection{Collaborative and Interdisciplinary Teaching Approaches}
\subsubsection{Cross-Disciplinary Projects}

As AI increasingly pervades all areas of technology, there is a growing need for interdisciplinary collaboration. Computer science courses can partner with departments such as data science, cybersecurity, or even digital humanities to create projects that leverage AI tools. Such projects can provide students with a broader perspective on how AI is applied in various domains and foster skills that are transferable across disciplines.

\subsubsection{Integrating Ethics and Responsible Use}

Ethical considerations must be an integral part of any curriculum that incorporates generative AI. Courses that focus on the responsible use of AI should be developed, covering topics such as bias in training data, the potential for AI-generated misinformation, and issues related to intellectual property. This ensures that future professionals are not only technically competent but also ethically aware.

\section{Assessment Strategies: How Should We Assess in the AI Era?}

The traditional methods of assessing  skills—typically through take-home assignments or timed  tests—are increasingly challenged by the availability of generative AI. A recent experiment involved the author (not a great programmer by any measure) being given an MSc level 15 week project to produce a cloud based, secure document-sharing system: a version navigating the API of a commercial cloud provider with security protocols and an effective web interface was produced in less than 8 minutes with Generative AI support, and would easily have passed under previous assessment regimes. The production of large numbers of lines of code no longer represents a depth of understanding or effort.  In this section, we discuss innovative assessment approaches that can better capture a student’s true understanding and ability.

\subsection{Process-Oriented Assessments}

\subsubsection{Submission of Drafts and Development Logs}

One effective strategy is to require students to submit the entire development process along with the final artifact. This is currently done in many institutions for plagarism detection.  In programming one might require for example the git logs to demonstrate progression of code development; for design concepts, intermediate steps and annotated changes may be handed in.

This may include:
\textit{Drafts and Revision Histories:} By examining intermediate versions of a project, instructors can gauge a student’s learning trajectory.
\textit{Annotated Development Logs:} Logs that explain design decisions, discuss alternatives, and iterations help differentiate between AI-assisted work and independent problem-solving.
\textit{AI transcripts:} submitting the prompts and discussions with the AI tool can provide insights into developing understanding and highlight areas of the code that are under-examined.

\subsubsection{Oral Examinations and Code Walkthroughs}

Oral examinations and artifact walkthroughs provide an opportunity for instructors to probe a student’s understanding in real time. During these sessions, students are asked to explain their system (whether code, design, interactive system or other artifact), discuss alternatives, and justify their design decisions. This method is particularly effective at discerning whether the student truly understands the solution or merely relied on AI-generated content.

\subsection{Performance-Based Assessments}

\subsubsection{Live Coding Sessions}

In-class live coding sessions, where students must solve a problem on the spot, can be a valuable assessment tool. These sessions minimize the opportunity for external AI assistance and help instructors assess a student’s spontaneous problem-solving ability. While stressful, live coding tests encourage students to internalize programming concepts rather than rely on AI shortcuts.  Alternatively, live coding sessions can allow the use of the AI tool, with the student verbalising how they are using it and responding to its outputs.

These can be modified for use in other parts of the computing curriculum - a prototype system can be used as a basis for student critique, with out without GenAI help, and the creation of alternative designs, workflows,  information architectures or UI designs to be done there and then can demonstrate their mastery of the tools and knowledge of what ot apply when.

\subsubsection{Collaborative Projects and Peer Review}

Group projects where students collaborate on a coding challenge can also serve as effective assessments. Peer review processes, in which team members evaluate each other’s contributions, further ensure accountability and provide a more holistic view of each student’s abilities. Collaborative assessments simulate real-world  development environments and underscore the importance of teamwork and communication.

\subsection{Incorporating Adaptive and Reflective Components}
\subsubsection{Adaptive Testing Methods}

Adaptive testing—where the difficulty of subsequent tasks is adjusted based on the student’s performance—can help ensure that assessments accurately reflect a student’s understanding. For instance, a programming test might begin with simpler tasks and then gradually introduce more complex challenges if the student demonstrates proficiency. This method provides a more individualized assessment of competence.

\subsubsection{Reflective Essays and Self-Assessment}

Requiring students to reflect on their coding process and evaluate the role of AI in their work can deepen their understanding and provide insight into their learning. Self-assessment components, where students critique their own work and outline areas for improvement, encourage a growth mindset and foster lifelong learning.  However, these too are subject to LLM generation, and so should be used only as one component in any assessment.

\subsection{Preparing for the Future: Emerging Topics and Interdisciplinary Skills}
Beyond reinforcing traditional skills and AI literacy, computer science education must also evolve to prepare students for emerging challenges:

\textit{Human-AI Collaboration:}
Courses should focus on how to effectively collaborate with AI systems. This might include project-based learning where students work in teams that include AI “partners” to solve complex problems.
\textit{Software Verification and Validation:}
As AI-generated code becomes more common, new methodologies for verifying and validating such code will be essential. Students should be exposed to both classical testing methods and emerging techniques that leverage AI for automated testing.
\textit{Interdisciplinary Approaches:}
Many modern CS problems are interdisciplinary. Integrating topics such as data ethics, cybersecurity, and human–computer interaction can provide students with a more holistic view of the challenges they will face in industry.
\textit{Lifelong Learning and Adaptability:}
Given the rapid pace of technological change, curricula should emphasize the importance of lifelong learning. Teaching students how to learn independently—by using AI as a tool for continuous professional development—will be critical.  It should be particularly noted that the pace of LLM change is almost unprecedented - in researching this article, the author notes that studies from 2023 are generally less positive about the roles of LLMs and highlight errors and issues that have often been eradicated by 2024 or 2025.

\section{Policy Recommendations for Computer Science Departments}

In light of the opportunities and challenges discussed above, computer science departments must implement policies that balance the benefits of generative AI with the need to maintain academic integrity and rigorous learning standards. The following policy recommendations are aimed at achieving this balance.

\subsection{Redesign Assessments to Emphasize the Learning Process}
\begin{enumerate}
\item \textit{Vivas:} Include oral examinations and in-class tests to demonstrate  understanding in real time.  Provide for vivas for more in-depth examination of larger credit work.
\item \textit{Adopt Collaborative Assessment:} Utilize adaptive testing methods and  collaborative projects that reflect real-world software development practices.
\item \textit{Process Documentation:} Mandate the submission of development logs, drafts, and artifact walkthrough recordings.
\end{enumerate}

\subsection{Enhance Faculty and Student Training}
\begin{enumerate}
    \item \textit{Professional Development Workshops:} Organize  workshops for faculty on best practices.
    \item \textit{AI Literacy Programs for Students:} Incorporate modules on AI ethics, critical evaluation of AI, and responsible use of generative AI into courses.
    \item \textit{Interdisciplinary Forums:} Establish interdisciplinary committees to continuously refine policies as AI technology evolves.
    \item \textit{Student voices:} ensure student perspectives are understood and taken into account.
\end{enumerate}

\subsection{Establish Clear Guidelines on AI Use}
\begin{enumerate}
    \item \textit{Define Acceptable Practices:} Clearly delineate if any uses of AI are unacceptable.
    \item \textit{Mandatory Disclosure:} Require students and faculty to disclose use of AI tools, including honour codes.
    \item \textit{Illustrative Documentation:} Provide concrete examples of acceptable and unacceptable uses of AI.

\end{enumerate}

\section{Discussion: Reconciling Innovation and Integrity in CS Education}

\subsection{Balancing Efficiency with Deep Learning}
The integration of generative AI in computer science education is undeniably transforming the learning process. AI-powered  assistants can reduce the time spent on routine tasks, allowing students to focus on higher-order thinking and complex problem solving. However, if these tools are used as shortcuts rather than learning aids, there is a risk that students will not develop the deep, conceptual understanding that is essential for long-term success.

Educators must therefore strike a balance. While it is beneficial to incorporate AI to foster efficiency, it is equally important to design curricula and assessments that require students to engage in critical reflection, debugging, and iterative development. By emphasizing the process of learning rather than just the final product, educators can help ensure that the use of AI supplements rather than supplants fundamental human, design, coding and development skills.

\subsection{The Role of Institutional Culture and Leadership}
Successful integration of AI into CS education depends not only on technical tools but also on a supportive institutional culture. Departmental leadership must actively promote ethical practices, ensure equitable access to AI resources, and foster continuous dialogue about best practices. By investing in training and establishing clear policies, institutions can create an environment in which both faculty and students are empowered to use AI responsibly.

\subsection{Future Research Directions}
While the current literature provides a solid foundation, further research is needed to assess the long-Term impact: longitudinal studies are required to evaluate how AI-assisted learning affects student outcomes over time.  We must explore new pedagogical models to foster innovative teaching strategies that integrate AI in collaborative, interdisciplinary contexts: these need creation, and should be rigorously tested and refined.  In addition, more data is needed on how AI usage differs across demographic groups and how educational institutions can mitigate any resulting disparities.

\section{Conclusion}

Generative AI is reshaping the field of computer science education in profound ways. With LLMs capable of writing and debugging code, emulating users, creating prototypes or evaluating systems, the traditional paradigms of teaching, learning, and assessment are undergoing rapid transformation. While these technologies offer tremendous opportunities to enhance learning efficiency and foster innovation, they also present serious challenges.

This article has examined the opportunities for integrating AI into CS education—from code generation and debugging to personalized feedback and collaborative learning—while also highlighting the risks of over-reliance, plagiarism, and superficial learning. We have discussed what should be taught in the AI era, including a renewed emphasis on core programming fundamentals, AI literacy, ethical considerations, and interdisciplinary skills. Moreover, we have explored innovative pedagogical strategies such as AI-augmented pair programming, flipped classroom models, and process-oriented assessments that aim to assess not only the final output but also the learning process.

To navigate these challenges, robust institutional policies are essential. Redesigned assessments that emphasize the process of coding are imperative; enhanced training for both faculty and students are all critical components of a successful policy framework. Institutional leadership and a culture that values both innovation and integrity will be key to harnessing the potential of generative AI without compromising educational quality.

In conclusion, computer science education stands at a crossroads. The decisions made today regarding curriculum design, teaching methods, and assessment practices will shape the future of the field. By embracing the opportunities offered by generative AI and implementing thoughtful, proactive policies, educators can prepare students not only to work effectively with AI tools but also to become creative, critical thinkers in an increasingly automated world.

\section*{Acknowledgements}

The work of many researchers, students and educators informed this article and are referenced appropriately.  Prompt engineering with GenAI (ChatGPT o3-mini-high, 4.0, 4.5, Claude Sonnet 3.5, 3.7, Google Gemini) were used for both the exploratory programming experiments alluded to in the text, and for searching for and identifying key academic work.  This was undertaken via extensive conversational refinement and  modifications of context and focus.   Any text produced by the LLM has been incorporated into existing copy, revised, rewritten and formulated according to the author's themes and ideas.

\bibliographystyle{agsm}
\bibliography{references}

\end{document}